\documentclass{moriond}

\def\be{\begin{equation}}
\def\ee{\end{equation}}
\def\bea{\begin{eqnarray}}
\def\eea{\end{eqnarray}}

\newcommand{\Photo}{}

\begin{document}
\vspace*{4cm}
\title{Open bottom production at NNLO+NNLL}

\author{Terry Generet}

\address{Cavendish Laboratory, University of Cambridge, Cambridge CB3 0HE, United Kingdom}

\maketitle\abstracts{
In this talk, I presented some of the results of the first calculation of open bottom production at hadron colliders at NNLO+NNLL, i.e.~a next-to-next-to-leading-order calculation that resums collinear logarithms at next-to-next-to-leading-logarithmic accuracy. This new computation achieves significantly reduced theory errors compared to previous calculations, with errors of just a few percent at high transverse momenta. These results are compared to data from many measurements performed at the Tevatron, where lower-order predictions have previously been found to underestimate the cross section. To perform such comparisons, the hadronisation and decay of the $b$-quark are included in the theory calculation where needed, yielding predictions for a wide range of final states.}

\section{Introduction}

There was a long history of excesses in measurements of the $b$-quark production cross section at the Tevatron\footnote{See e.g.~the work of Cacciari and Nason\cite{Cacciari:2002pa} for a brief overview.}. Initial comparisons to theory predictions\cite{Nason:1987xz,Nason:1989zy,Beenakker:1990maa} at next-to-leading order (NLO) in QCD suggested the measured cross section to be a factor of three larger than predicted. This was later reduced to a factor $\sim1.7$ by Cacciari and Nason\cite{Cacciari:2002pa}, by including large transverse-momentum logarithms to all orders in perturbation theory at next-to-leading-logarithmic accuracy (NLL) and by improving the non-perturbative model used to describe the hadronisation of $b$-quarks to $B$-hadrons. While this was a significant improvement, the agreement between theory and experiment was still marginal. It would take almost 20 years before the calculation was improved further: the next-to-next-to-leading order (NNLO) corrections to massive $b$-quark production at hadron colliders were presented for the first time by Catani \textit{et al.}\cite{Catani:2020kkl} In that work, a comparison\footnote{Differential predictions were provided for both the Tevatron and the LHC. However, only the LHC predictions were compared to data at the differential level.} to a single measurement of the fiducial cross section at the Tevatron\cite{CDF:2004jtw} showed that the inclusion of NNLO corrections significantly improves the agreement with data, providing a first confirmation that higher-order effects significantly affect the data/theory comparison and have the potential to explain the apparent Tevatron excess entirely.

The goals of the work presented here are as follows. First, an independent NNLO computation is compared to many different measurements made at the Tevatron and the S$p\overline{p}$S. This includes both differential distributions and fiducial cross sections. The second goal is to finally extend the NLO+NLL calculation, specifically as implemented in the FONLL scheme\cite{Cacciari:1998it,Cacciari:2001td}, to NNLO+NNLL. The technical details of these calculations are not covered here. They will be published, together with the (preliminary) results presented in this talk, in a future publication\cite{CGMP}. The calculations are based on the \textsc{Stripper} framework\cite{Czakon:2010td,Czakon:2011ve,Czakon:2014oma,Czakon:2019tmo} and involve an extended version of its implementation of fragmentation\cite{Czakon:2021ohs,Czakon:2022pyz}.

\section{Results}

\begin{figure}
\includegraphics[width=0.49\linewidth]{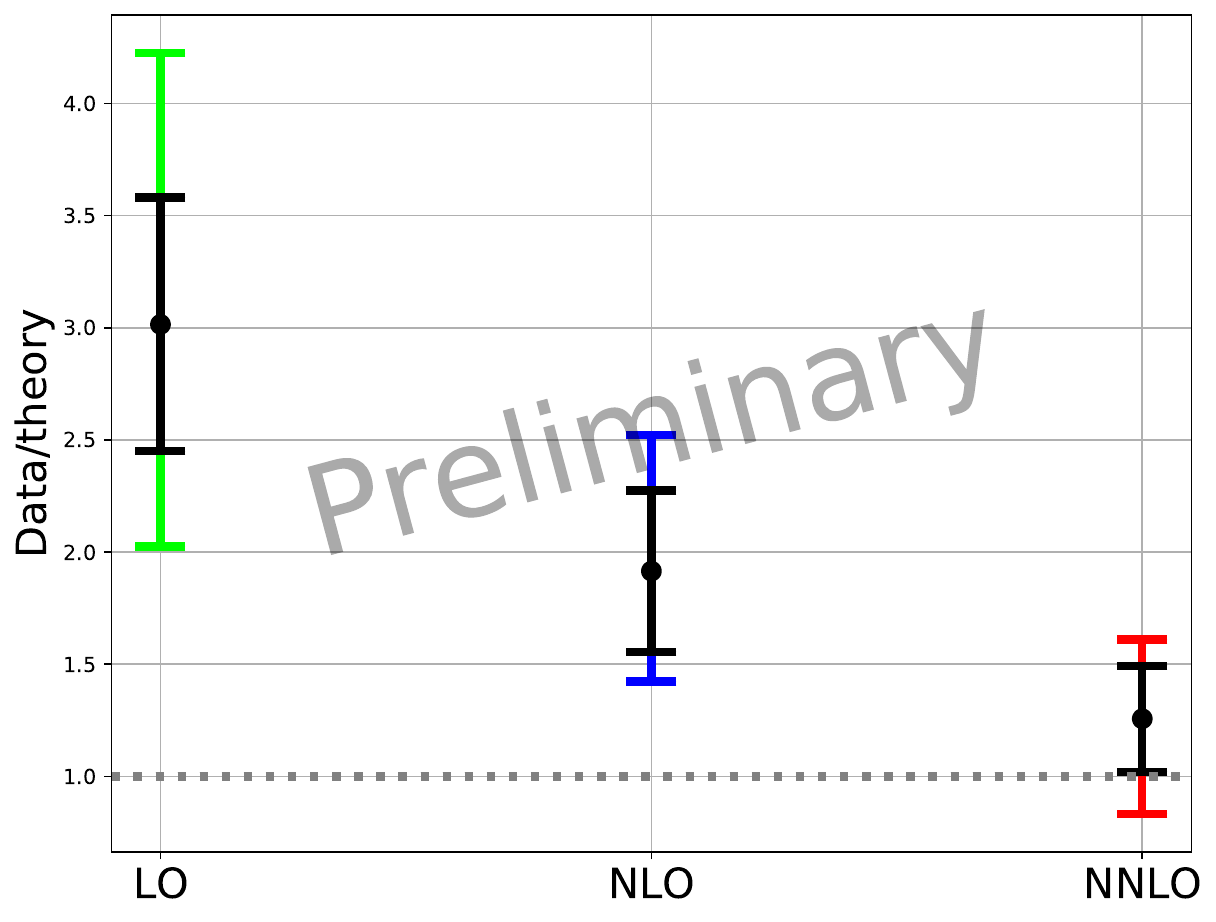}
\includegraphics[width=0.49\linewidth]{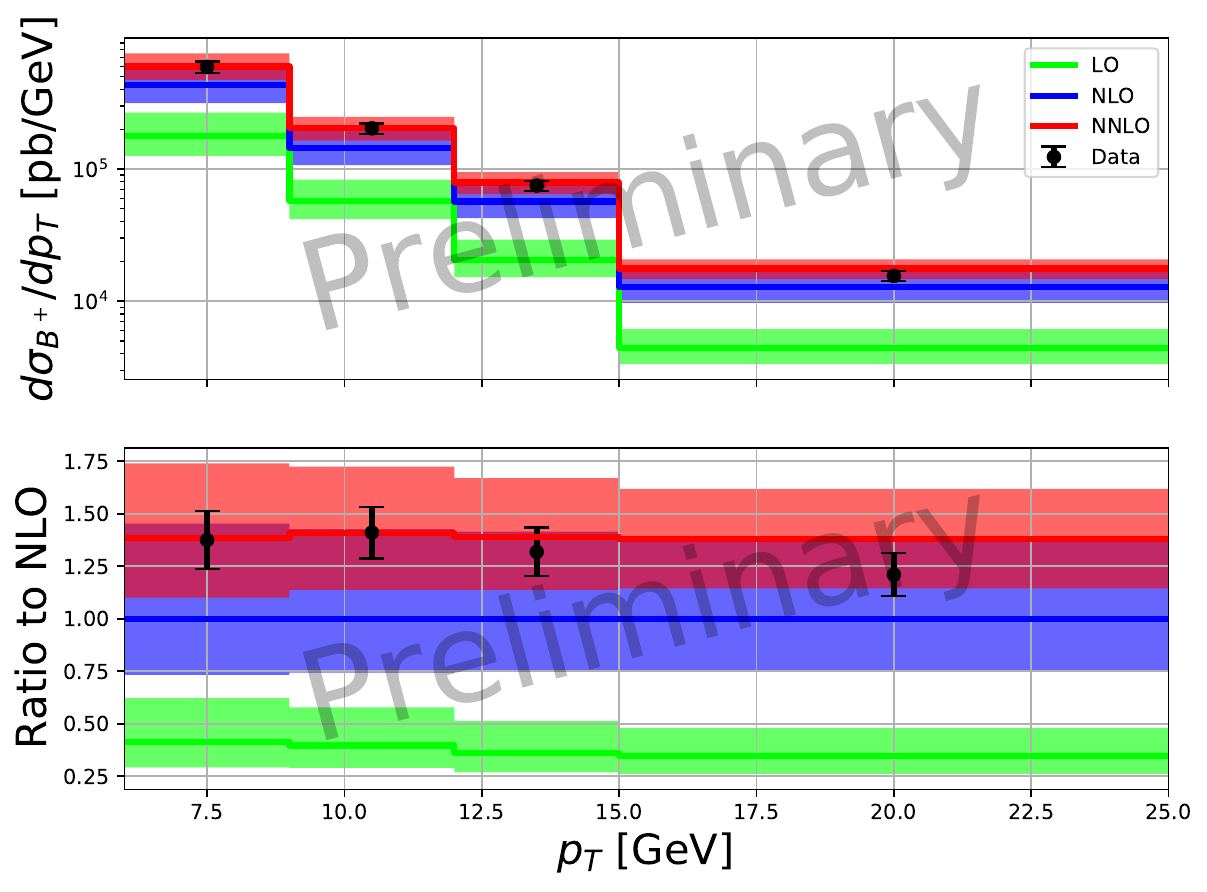}
\caption[]{Left: The average of the data-over-theory ratio at LO, NLO and NNLO. The coloured errors show the standard 7-point scale uncertainties, while the black errors represent the experimental uncertainties. Right: A comparison of the predicted $B$-hadron $p_T$-spectrum through NNLO with the CDF data\cite{CDF:2006ipg}. The theory uncertainties again correspond to scale uncertainties, while the experimental ones are obtained by adding statistical and systematic uncertainties in quadrature.}
\label{fig:TevatronRatio}
\end{figure}

The first result is an average of the data/theory ratio of many different measurements of the fiducial $b$-quark production cross section\cite{UA1:1985bnq,UA1:1990vvp,CDF:1992sue,CDF:1993dog,CDF:1993fax,UA1:1993jok,CDF:1994zxd,CDF:1996kan,D0:1999hay,CDF:2007pxr}. In practice, these are measurements of different final states (fully or partially reconstructed $B$-hadrons, in different decay channels), unfolded to the $b$-quark level by the respective experimental collaborations. These include measurements of single-inclusive $b$-quark production, as well as correlated production of $b\overline{b}$-pairs, with $b$-quark-transverse-momentum cuts ranging from 5 GeV to 54 GeV. It was checked that the data/theory ratio does not show a significant dependence on the transverse momentum cut. The results are shown in the left panel of figure \ref{fig:TevatronRatio}. The leading-order (LO) prediction is off by a factor of three. The NLO corrections reduce this discrepancy to a factor of two. However, the ratio is still inconsistent with unity within either the experimental or the theory uncertainties. The NNLO corrections further reduce the apparent excess to just a factor $\sim1.3$, which is not only significantly lower, but also consistent with one within both the experimental and theory uncertainties. It would therefore appear that this long-standing discrepancy has finally been resolved.

The right panel of figure \ref{fig:TevatronRatio} shows a comparison of theory predictions with a CDF measurement of the $B^+$ $p_T$-spectrum\cite{CDF:2006ipg}. A similar conclusion can be drawn here: while the NLO predictions are only marginally consistent with the data within the large scale uncertainties, the NNLO predictions agree almost perfectly. Note that, while the scale uncertainties are reduced when including the NNLO corrections, the reduction is not particularly impressive. This is, of course, to be expected, since the relevant scale for this process is the $b$-quark mass, which, while perturbative, is a rather low scale. Therefore, the strong coupling is large, and the perturbative series converges slowly.

\begin{figure}
\includegraphics[width=0.49\linewidth]{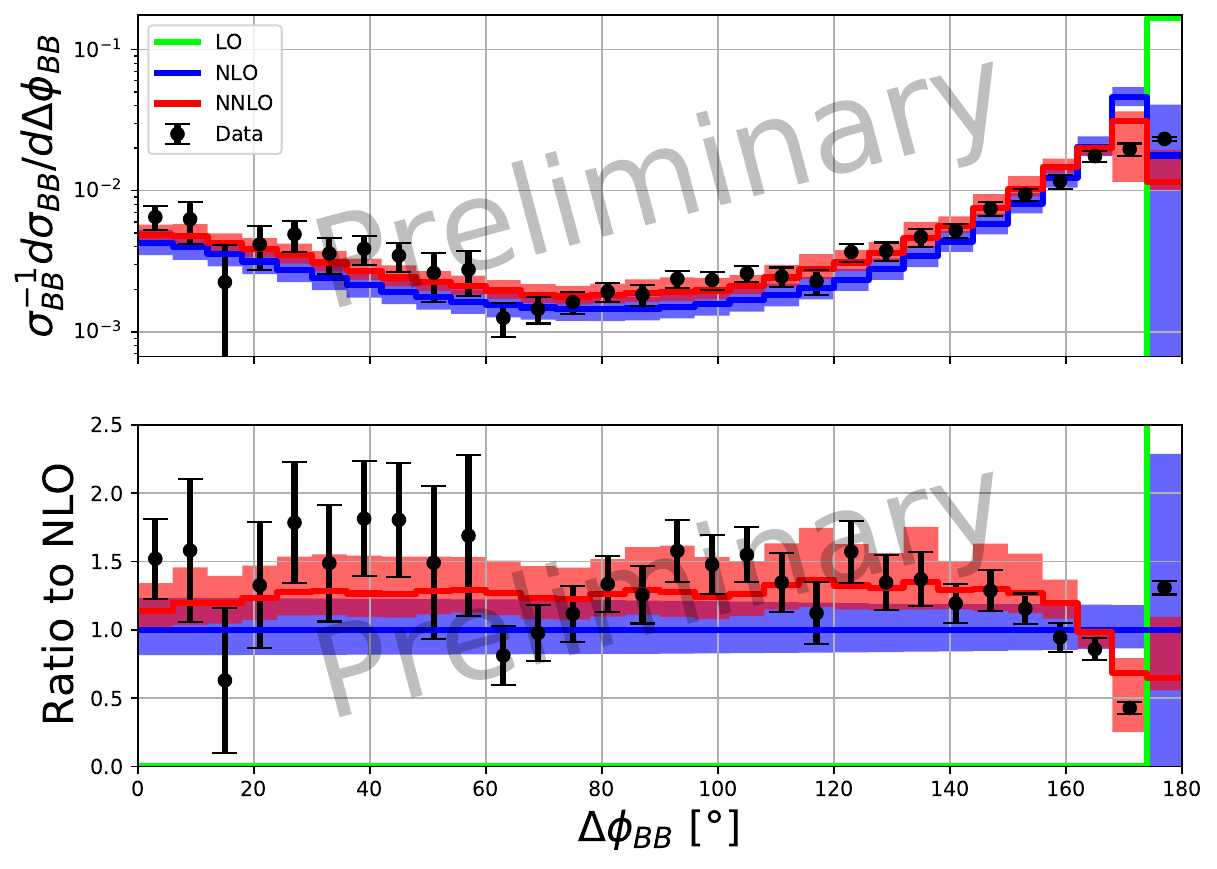}
\includegraphics[width=0.49\linewidth]{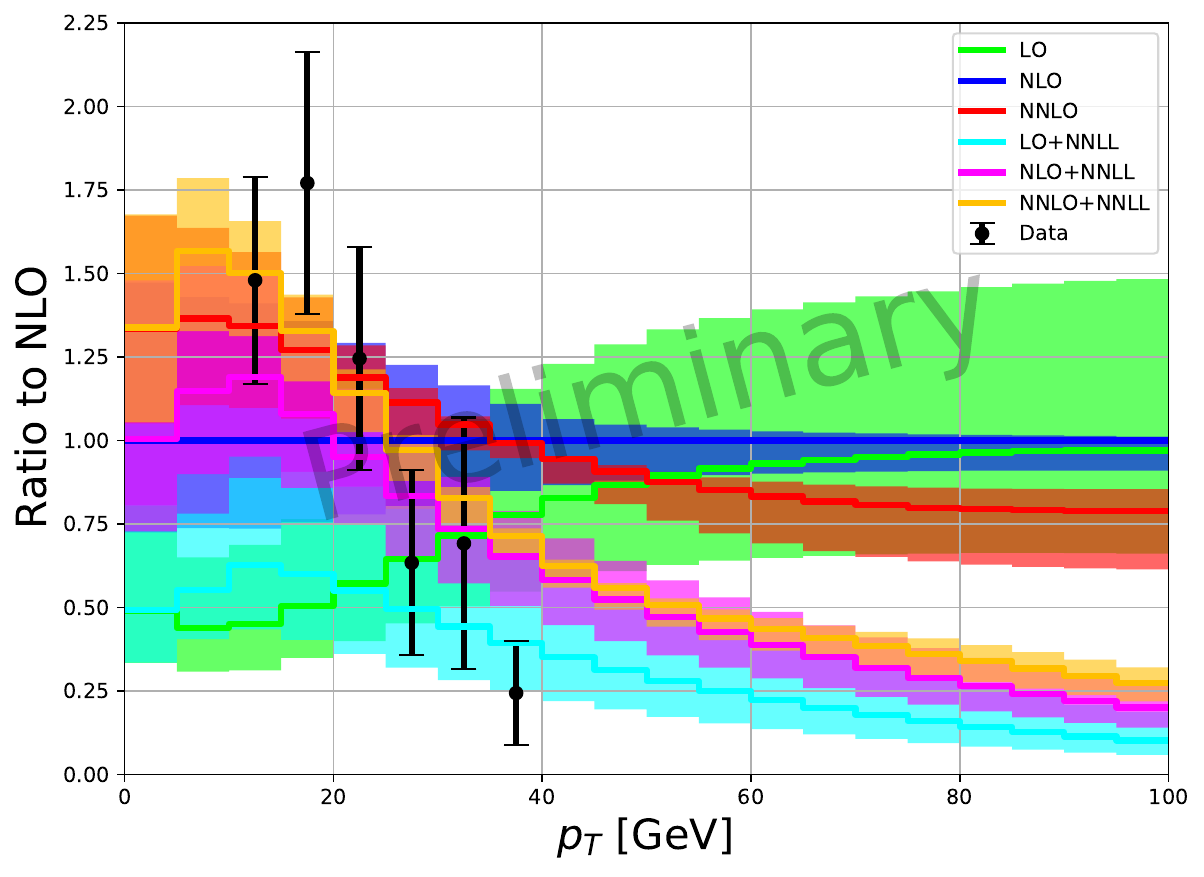}
\caption[]{Left: Distribution of the azimuthal angle between two $B$-hadrons through NNLO compared to CDF data\cite{CDF:2004mmv}. Right: The $p_T$-spectrum of muon from $B$-hadron decays as measured by UA1\cite{UA1:1990vvp} compared to theory predictions at various levels of perturbative accuracy. Unlike all other theory predictions presented here, the scale uncertainties on the resummed predictions are estimated using a 15-point scale variation.}
\label{fig:B_angular}
\end{figure}

The effects of higher-order corrections on the angular distribution of $B$-hadrons can be seen in the left panel of figure \ref{fig:B_angular}, where fixed-order predictions through NNLO are compared to data measured by CDF\cite{CDF:2004mmv}. There is a clear shape difference between the measured distribution and the one obtained using the NLO computation. The NNLO corrections essentially perfectly account for this shape difference, significantly improving the description of this observable, which is highly sensitive to the details of QCD radiation.

The right panel of figure \ref{fig:B_angular} shows the first NNLO+NNLL result for open bottom production at a hadron collider. Specifically, it shows the transverse-momentum distribution of muons originating from $B$-hadron decays at the S$p\overline{p}$S through NNLO, both with and without NNLL resummation of collinear logs. Also shown are the data points measured by the UA1 collaboration\cite{UA1:1990vvp}. Focusing on just the fixed-order predictions, it is clear that perturbation theory breaks down at large transverse momenta: the NNLO curve lies well outside the NLO scale band and the NNLO scale band is significantly wider than the NLO one. This is to be contrasted with the resummed results, which show excellent convergence: there is significant overlap between the scale bands of consecutive perturbative orders and the width of the scale band is consistently reduced as the order is increased. The resummed predictions also follow the measured distribution much more closely. The NNLO+NNLL scale uncertainties are significantly smaller than the NLO+NNLL ones, indicating a significant improvement in precision compared to the previous standard of NLO+NLL\footnote{While no NLO+NLL predictions are shown here, it was checked that the difference between NLO+NLL and NLO+NNLL is minor, both in terms of their central values and in terms of precision.}.

\section{Conclusion and outlook}

I have presented comparisons of proton-anti-proton-collider data on open bottom production to cutting-edge NNLO predictions. The inclusion of the NNLO corrections significantly improves the agreement with the data, both in terms of normalisation, solving a long-standing discrepancy with Tevatron data, and in terms of the shape of certain distributions. I have also presented the first results at NNLO+NNLL accuracy, which indicate a significant improvement in precision compared to the best predictions previously available. In the near future, these calculations will be used in an analysis of LHC data, which is significantly more precise than previously available predictions. As such, the increased precision of the new predictions will be of great value in restoring parity between experimental and theory uncertainties in open bottom production at hadron colliders.

\section*{Acknowledgments}

I would like to thank the organisers of the 58th Rencontres de Moriond for giving me the opportunity to present this work. It is a truly unique conference made possible by their organisational efforts. This work has been supported by STFC consolidated grants ST/T000694/1 and ST/X000664/1.

\end{document}